\documentclass[aps,prl,twocolumn]{revtex4-1}
\usepackage{graphicx}
\usepackage{amsmath}
\usepackage{dcolumn}
\usepackage{bm}
\usepackage{float}
\usepackage{epstopdf}

\begin{document}

\preprint{}

\title{Broadband spectroscopy of thermodynamic magnetization fluctuations through a ferromagnetic spin-reorientation transition}

\author{A. L. Balk$^{1}$, F. Li$^{2,3}$, I. Gilbert$^{4}$, J. Unguris$^4$, N. A. Sinitsyn$^{2}$, S. A. Crooker$^1$}

\affiliation{$^1$National High Magnetic Field Laboratory, Los Alamos National Laboratory, Los Alamos, NM 87545, USA}
\affiliation{$^2$Theoretical Division, Los Alamos National Laboratory, Los Alamos, NM 87545, USA}
\affiliation{$^3$Center for Nonlinear Studies, Los Alamos National Laboratory, Los Alamos, NM 87545, USA}
\affiliation{$^4$Center for Nanoscale Science and Technology, National Institute of Standards and Technology, Gaithersburg, MD 20899, USA}

\begin{abstract}
We use scanning optical magnetometry to study the broadband frequency spectra of spontaneous magnetization fluctuations, or ``magnetization noise", in an archetypal ferromagnetic film that can be smoothly tuned through a spin reorientation transition (SRT). The SRT is achieved by laterally varying the magnetic anisotropy across an ultrathin Pt/Co/Pt trilayer, from the perpendicular to in-plane direction, via graded Ar$^+$ irradiation.  In regions exhibiting perpendicular anisotropy, the power spectrum of the magnetization noise, $S(\nu)$, exhibits a remarkably robust $\nu^{-3/2}$ power law over frequencies $\nu$ from 1~kHz to 1~MHz. As the SRT region is traversed, however, $S(\nu)$ spectra develop a steadily-increasing critical frequency, $\nu_0$, below which the noise power is spectrally flat, indicating an evolving low-frequency cutoff for magnetization fluctuations.  The magnetization noise depends strongly on applied in- and out-of-plane magnetic fields, revealing local anisotropies and also a field-induced emergence of fluctuations in otherwise stable ferromagnetic films. Finally, we demonstrate that higher-order correlators can be computed from the noise.  These results highlight broadband spectroscopy of thermodynamic fluctuations as a powerful tool to characterize the interplay between thermal and magnetic energy scales, and as a means of characterizing phase transitions in ferromagnets. 
\end{abstract}

\maketitle

Spontaneous magnetization fluctuations can occur in ferromagnetic materials, even in thermal equilibrium, particularly when the magnetic anisotropy energy becomes comparable to or less than the available thermal energy \cite{Bittel, Ocio, Silva, Smith, WeissmanSG, Bean, Sinwani}. These intrinsic thermodynamic fluctuations encode valuable information about the magnetization dynamics of the system itself, because their frequency spectrum $S(\nu)$ is intimately and necessarily related to the dissipative (imaginary) part of the magnetic susceptibility $\chi''(\nu)$, in accord with the fluctuation-dissipation theorem \cite{Kubo, Ocio} (namely, $\chi''(\nu) \sim \frac{\nu}{k_B T}S(\nu)$, where $k_BT$ is the thermal energy). Spectroscopy of this intrinsic ``magnetization noise" can therefore provide an alternative and entirely passive means of measuring magnetization dynamics, that does not require driving, exciting, or perturbing the system away from thermal equilibrium -- in contrast to most conventional methods for measuring magnetic resonance or ac susceptibility.

A fascinating and technologically-relevant testbed in which to study thermodynamic magnetization fluctuations are thin ferromagnetic films, because magnetic anisotropy energies can be readily engineered over a very wide range via material choice and by growth and post-processing conditions \cite{Johnson, Vaz}. This tunability arises from the competition and delicate balance between shape anisotropy (which generally favors in-plane magnetic alignment) and interfacial anisotropy (which can favor out-of-plane magnetic alignment for certain material combinations).  Especially interesting are films in which the total magnetic anisotropy is continuously tuned to and through zero, by laterally varying either the film's thickness \cite{Allenspach, Qiu} or its degree of interfacial disorder \cite{Chappert, Fassbender}. In this case the film can exhibit a spin-reorientation transition (SRT) \cite{Pappas, Pescia, Jensen, Millev} wherein the direction of ferromagnetic ordering transitions from out-of-plane (perpendicular) to in-plane. In perpendicularly-magnetized films with small net anisotropy, it is well established that the combination of exchange and dipolar energies leads to the formation of maze-like patterns of magnetic domains \cite{Hubert, Yafet}, as observed and studied extensively in ultrathin films of (Fe/Ni)/Cu \cite{Wu, Won, Kronseder2012}, Co/Pt \cite{Belliard, Bergeard, Ando}, and CoFeB \cite{Yamanouchi}. As the SRT is approached and the net magnetic anisotropy is reduced to zero, it has recently been observed that these domain patterns begin to fluctuate markedly in time. These thermodynamic magnetization fluctuations have been measured by electron microscopy \cite{Portmann, Kronseder, Kronseder2012, Meier}, magneto-optic Kerr effect (MOKE) \cite{Bergeard, Balk}, x-ray scattering \cite{Seu}, and transport \cite{Diao}, revealing spatial correlations, topological effects, suceptibilities, and higher-order anisotropy. However, timescales of these measurements are typically slow, being limited to $>$1 ms. Fluctuations at frequency scales faster than 1~kHz were not resolved. However, the full frequency spectrum of these fluctuations contains rich information about the distributions of relaxation rates and the evolution of the anisotropy landscape through the SRT, which provide important insight for developing theoretical models.

To address this need and to complement these recent studies, here we develop a fast optical magnetometer to investigate the broadband frequency spectrum of thermodynamic magnetization fluctuations in ferromagnetic films that are smoothly tuned through a SRT. Using Pt/Co/Pt trilayers with laterally-graded magnetic anisotropy, we find that the frequency spectrum of the magnetization noise, $S(\nu)$, exhibits a remarkably robust $\nu^{-3/2}$ power law from 1~kHz to 1~MHz in regions exhibiting perpendicular magnetization, indicating a broad distribution of fluctuation rates.  However as the SRT region is traversed, $S(\nu)$ develops a steadily-increasing critical frequency ($\nu_0$) below which the noise tends towards spectrally flat, indicating a minimum relaxation rate that becomes increasingly fast through the SRT. The magnetization fluctuations also depend strongly on applied magnetic fields, which can be understood within the context of the fluctuation-dissipation theorem and from considerations of the magnetic free energy.  Field-dependent maps of the noise reveal detailed local magnetic anisotropies, and field-induced fluctuations are found to emerge even in trilayers with nominally stable perpendicular ferromagnetism.  Finally we show that these methods can be used to analyze higher-order noise correlations, which can probe non-gaussian noise and time-reversal breaking effects.  These results demonstrate that broadband optical detection of magnetization fluctuations provides a powerful and straightforward method for studying the subtle interplay between thermal energy and magnetic anisotropy that can drive phase transitions in ferromagnets, and in particular the spin-reorientation transition in ultrathin films.

\begin{figure*}[tbp]
\center
\includegraphics[width=0.85\textwidth]{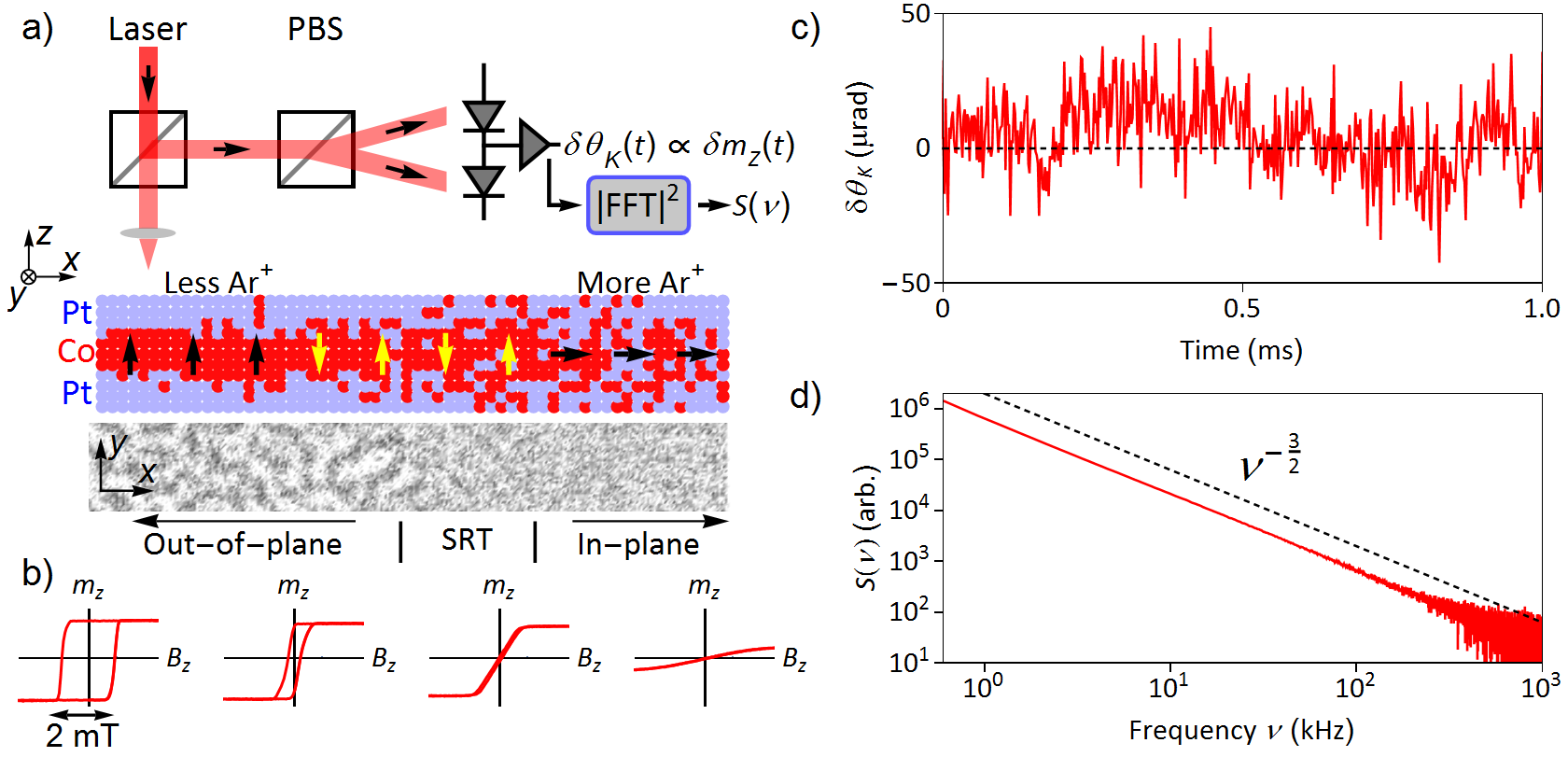}
\caption{a) Experimental schematic: A scanning broadband optical magnetometer based on polar MOKE measures the spontaneously fluctuating out-of-plane magnetization $\delta m_z(t)$ within a small (4~$\mu$m) spot on the sample.  The power spectral density of this magnetization noise, $S(\nu)$, is continuously computed and averaged in real time. The sample is a Pt/Co/Pt trilayer irradiated with a laterally-varying dose of Ar$^+$, generating a gradient of interfacial magnetic anisotropy.  Red and blue dots depict Co and Pt atoms; arrows depict the net magnetization direction, with yellow arrows indicating the region where the spin reorientation transition (SRT) occurs. The magnetic contrast image of the trilayer (adapted from \cite{Balk}) surface shows the maze-like domain patterns that form in graded trilayers of this type. The patterns become blurry near the SRT as fluctuations increase and occur faster than the image acquisition time. b) Conventional magnetization measurements $m_z(B_z)$ show that the coercive field $\mu_0 H_c \rightarrow 0$ as the SRT is approached. c) Characteristic Kerr rotation fluctuations, $\delta \theta_K(t)$, reveal the thermodynamic magnetization fluctuations $\delta m_z(t)$.  d) $S(\nu)$ falls as a $\nu^{-3/2}$ power law over the measured 1~kHz to 1~MHz range in regions of the trilayer exhibiting perpendicular magnetization. Discussion of measurement uncertainties in this and subsequent figures can be found in the Supplemental Material.}
\label{fig1}
\end{figure*}

\subsection{Pt/Co/Pt trilayers with laterally-graded magnetic anisotropy}
Figure 1a depicts the experiment and samples. The samples are Ta(3.8 nm)/Pt(3.9 nm)/Co(0.8 nm)/Pt(1.9 nm) ultrathin ferromagnetic films sputtered on Si substrates under high vacuum conditions. Conventional polar MOKE measurements of the average out-of-plane magnetization ($m_z$ versus $B_z$) confirm that the as-grown films have perpendicularly-oriented magnetization, with approximately square hysteresis loops and coercive fields $\mu_0 H_c \approx 2$ mT. To tune the magnitude and direction of the net magnetic anisotropy, we irradiate the samples with a laterally-varying dose of low-energy (100~eV) Ar$^+$, which generates disorder and reduces the interfacial (perpendicular) magnetic anisotropy \cite{Chappert}. The net magnetization direction that results from the competition between interfacial and shape anisotropy is therefore oriented out-of-plane ($\parallel \pm \hat{z}$) on regions of the sample with lower disorder, but switches to in-plane orientation ($\perp \hat{z}$) on sample regions with higher disorder, as depicted. Between these regions the spin reorientation transition occurs.

Shown below the cartoon of the trilayer is a magnetic contrast image  of the surface of a laterally-graded Pt/Co/Pt trilayer (obtained via MOKE), showing the maze-like magnetic domain patterns that exist at $B$=0 in its fully relaxed and demagnetized state.  Dark and bright domains correspond to out-of-plane magnetization oriented along $+\hat{z}$ and $-\hat{z}$, respectively.  As also observed in many prior studies of ultrathin ferromagnets \cite{Wu, Kronseder2012, Balk, Meier}, the characteristic size of the domains decreases as the SRT region is approached and the perpendicular magnetic anisotropy is reduced. Near the SRT, the images become increasingly blurry as thermodynamic fluctuations of the domain walls increase and become faster than the image acquisition rate (typically video-rate, or $\approx$30~Hz \cite{Balk, Meier, Bergeard}, but as fast as 1~kHz in recent work by Kronseder \cite{Kronseder}). Our goal is to measure the relevant timescales and frequency distributions of these magnetization fluctuations over a much broader range of timescales and deep into (and through) the SRT, where the temporal resolution of imaging techniques can be limited.

Figure 1b shows conventional polar MOKE measurements of $m_z(B_z)$ across the sample, confirming the graded magnetic anisotropy: The open hysteresis loops become narrower as perpendicular anisotropy is reduced, and the coercive field $\mu_0 H_c \rightarrow 0$ as the SRT is approached. Near the SRT, no hysteresis is observed and the low-field magnetization is effectively linear (constant susceptibility $dm_z/dB_z$), indicating that the trilayer is able to relax to an equilibrium maze domain pattern on a timescale faster than the (slow) sweep rate of the the applied field $B_z$. Beyond the SRT the magnetization direction switches to in-plane and $m_z$ is suppressed.

\subsection{Optical spectroscopy of magnetization fluctuations}
To detect and quantify the broadband magnetization fluctuations that exist in thermal equilibrium in these Pt/Co/Pt trilayers, we developed a real-time optical noise magnetometer based on polar MOKE. A linearly-polarized probe laser (632.8~nm, $\approx$1~mW) is  focused to a small (4~$\mu$m diameter) spot on the sample, which in turn is mounted on an \textit{x-y} positioning stage for lateral scanning. This arrangement exclusively measures stochastic magnetization fluctuations in the out-of-plane direction, $\delta m_z (t)$, which impart a fluctuating Kerr rotation $\delta \theta_K (t)$ on the reflected probe laser that is detected using balanced photodiodes. The resulting noise signal $\delta \theta_K (t) \propto \delta m_z(t)$ is continuously digitized, processed, and averaged in real time. Typically we compute and analyze $S(\nu)$, the power spectral density of the fluctuation signal.  That is, $S(\nu) = \langle a(\nu) a^*(\nu) \rangle$ where $a(\nu)$ is the Fourier transform of $\delta m_z(t)$ and the brackets indicate an average over repeated measurements.  Equivalently, $S(\nu)$ is the real Fourier transform of the second-order (two-point) time correlator $\langle \delta m_z(0) \delta m_z(t) \rangle$. Importantly, we note that this setup also allows for measurement and analysis of higher-than-second order noise correlators, as demonstrated in the last section.

The measurement bandwidth is determined by the speed of the photodetectors and digitizers. For these Pt/Co/Pt trilayers we use low-noise detectors and amplifiers with frequency response of several megahertz. The spectral density of the measured Kerr rotation is small, typically below 1~$\mu$rad/$\sqrt{\textrm{Hz}}$. The fundamental photon shot noise of the probe laser itself contributes $\sim$50~nrad/$\sqrt{\textrm{Hz}}$ of (white) background noise; this constitutes the dominant source of non-magnetic noise at frequencies above a few hundred hertz. This shot noise can be mitigated by signal averaging and background subtraction.  At lower frequencies,  acoustic noise and mechanical vibrations limit accurate recovery of small magnetization fluctuations.  Unless otherwise noted, we therefore restrict detailed analysis to frequencies $\nu > 1$~kHz. These methods are adapted from studies of ``optical spin noise spectroscopy" in atomic vapors and semiconductors, which are typically based on Faraday rotation at much higher (MHz) frequencies \cite{Crooker, Oestreich, Crooker2, Zapasskii}.

Figure 1c shows an example of the spontaneous magnetization fluctuations $\delta m_z(t)$ that are measured in thermal equilibrium at an out-of-plane region of the Pt/Co/Pt trilayer, at room temperature ($T$=295~K) and without any applied magnetic field.  The induced Kerr rotation signal, $\delta \theta_K (t)$, fluctuates in time about zero with a typical standard deviation of several microradians. Figure 1d shows the corresponding power spectral density of this magnetization noise, $S(\nu)$, averaged over many minutes.  Strikingly, $S(\nu)$ exhibits a very robust power-law decay over the frequency range from $\nu$=1~kHz to 1~MHz, falling as $\nu^{-\alpha}$ where $\alpha \simeq 3/2$. As shown below, this power-law exponent is very robust and is independent of changes in magnetic anisotropy, temperature, and small applied magnetic fields $B$.

In general, power law dynamics indicate that a system cannot be characterized by a single timescale but rather exhibits a broad distribution of relaxation and fluctuation timescales \cite{Press, Weissman, Milotti}.  Given that many magnetic domains and fluctuating domain walls are simultaneously probed within the focused laser spot, it is therefore worthwhile to ask what distribution of timescales or relaxation rates might generate the measured $S(\nu) \sim \nu^{-3/2}$ noise spectrum.

In a simple case, consider a system comprising many independent simple fluctuators, each characterized by a single exponentially-decaying correlation time $\tau_i$ (i.e., $\langle m(0) m(t) \rangle \sim e^{-t/\tau_i}$).  Individually, each fluctuator therefore contributes a Lorentzian power spectral density, $S_i(\nu) \sim 1/(\nu^2 + \gamma_i^2)$, to the total measured noise spectrum; that is, $S_i(\nu)$ is approximately flat up to the characteristic relaxation rate $\gamma_i = \tau_i^{-1}$ before falling off as $\nu^{-2}$ at high frequencies. In this case the total noise spectrum is the weighted sum of many Lorentzians \cite{Milotti}, where the weight is determined by the distribution, $D(\gamma)$, of fluctuators having characteristic relaxation rate $\gamma$:
\begin{equation}
S(\nu) \propto \sum_i \frac{1}{\nu^2 + \gamma_i^2} \approx \int_{\gamma_{\rm min}} ^{\gamma_{\rm max}} \frac{D(\gamma)}{\nu^2 + \gamma^2} d\gamma.
\end{equation}
The distribution $D(\gamma)$ determines the functional form of $S(\nu)$, and the upper and lower cutoffs prevent  divergences. If $D(\gamma)$ itself follows a power-law $D(\gamma) \propto \gamma^{-\beta}$, then $S(\nu) \approx \nu^{-(\beta + 1)}$ for $\gamma_{\rm min} \ll \nu \ll \gamma_{\rm max}$.  Note that if $D(\gamma)$ is truncated below some minimum relaxation rate $\gamma_{\rm min}$, then $S(\nu)$ will deviate from a pure power law and will tend towards spectrally flat at low frequencies below $\gamma_{\rm min}$, as will become relevant in the next section. Moreover, if $D(\gamma)$ is truncated above a maximum rate $\gamma_{\rm max}$, then $S(\nu)$ will eventually decay as $\nu^{-2}$ at high frequencies. Therefore the $\nu^{-3/2}$ noise spectrum measured in Fig. 1d is (at least) consistent with an ensemble of simple fluctuators with distribution $D(\gamma) \propto \gamma^{-1/2}$. Furthermore, no signature of $\gamma_{\rm min}$ or $\gamma_{\rm max}$ is evident in the data -- at this location on the sample these limits clearly lie well below and above our measured frequency range, respectively. While clearly an oversimplification --ferromagnets are, after all, correlated systems and are not composed of independent fluctuators-- these arguments help to provide a basis for understanding how changes in the functional form of $S(\nu)$ can emerge.

A broad distribution of relaxation rates can be expected in ultrathin ferromagnetic films due to microscopic variations of the magnetic exchange, anisotropy, and dipolar energies. In particular, these variations can be caused by a spatially disordered magnetic energy landscape arising from the interface roughness, which can be comparable to film thickness. Indeed, a disordered magnetic energy forms the basis for a variety of statistical models \cite{Alessandro, Cizeau, Kuntz} which have been applied to understand, e.g., the Barkhausen (magnetic switching) noise that arises in ferromagnets that are driven by a changing applied magnetic field.  Predictions and measurements \cite{Narayan, Bohn} of the Barkhausen noise spectrum often yield power laws with exponents in the range of $-1.5$ to $-2$, in potential correspondence with the value of $\alpha$ determined in our studies of purely thermodynamic (undriven) magnetization fluctuations.

\subsection{Laterally scanning through the SRT at fixed temperature}
Having established a $\nu^{-3/2}$ power-law spectrum of magnetic fluctuations in a region of the trilayer exhibiting out-of-plane ferromagnetism, we now explore the much more interesting question of how the fluctuation spectrum evolves when the net magnetic anisotropy is reduced through zero and the sample undergoes a SRT. Figure 2 shows $S(\nu)$ as the probe laser is scanned laterally across the sample, beginning (on the left) at a region of low interfacial disorder and strong perpendicular magnetic anisotropy, then moving through regions of increasing disorder where the sample undergoes a SRT, and ending (on the right) in the region of large disorder where the magnetization orientation is in-plane.  As discussed in the previous section, Fig. 2a  shows that $S(\nu) \sim \nu^{-3/2}$ in regions of perpendicular magnetization. As the probe laser is scanned toward the SRT region and perpendicular magnetic anisotropy becomes weaker, $S(\nu)$ continues to exhibit a robust $\nu^{-3/2}$ power law.  However, the magnitude of $S(\nu)$ and therefore also the integrated noise power, $P = \int S(\nu) d\nu$, increases significantly.  Given that the thermal energy $k_B T$ is not changing, the increase in thermodynamic fluctuations is consistent with the reduction of magnetic anisotropy as the SRT is approached.

\begin{figure}[tbp]
\center
\includegraphics[width=.45\textwidth]{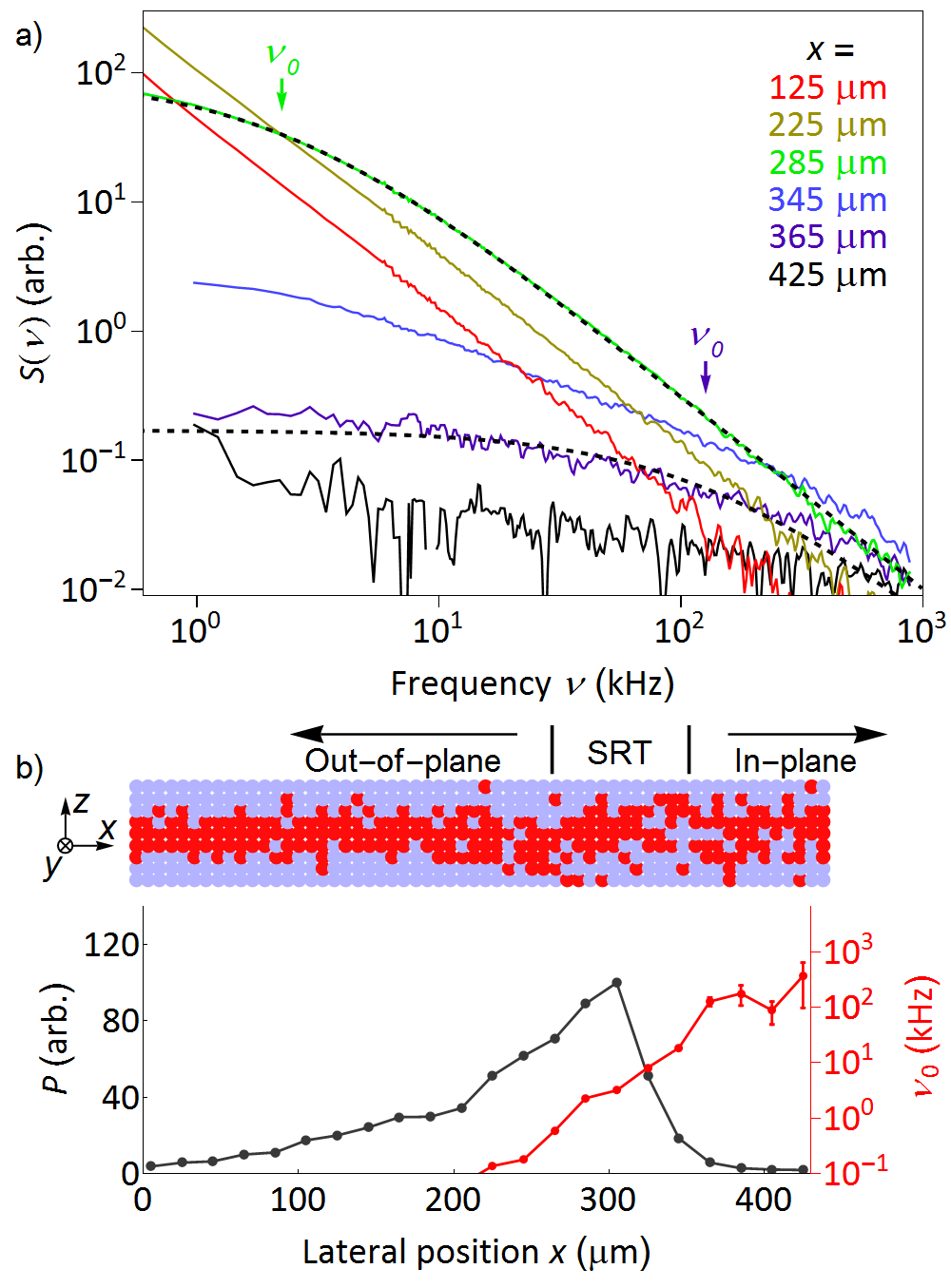}
\caption{a) Power spectra of the magnetization fluctuations, $S(\nu)$, at different lateral positions $x$ on the Pt/Co/Pt trilayer. In regions with perpendicular magnetization, $S(\nu) \propto \nu^{-3/2}$. Upon entering and traversing the SRT region, $S(\nu)$ develops a steadily increasing critical frequency, $\nu_0$, below which the noise tends toward spectrally flat, consistent with an evolving low-frequency cutoff in the distribution of fluctuation rates. Empirically, all spectra can be fit by $S(\nu) \propto (\nu+\nu_{0})^{-3/2}$ (dashed lines). b) The integrated noise power, $P = \int S(\nu) d\nu$, exhibits a peak in the SRT region.  Concomitantly, $\nu_0$ rapidly increases through the measurement bandwidth (red points).}
\label{fig2}
\end{figure}

Crucially, the functional form of the magnetization fluctuation spectrum begins to evolve dramatically upon entering the SRT region. As seen in Fig. 2a, $S(\nu)$ spectra develop a critical frequency, $\nu_0$, below which the noise power rolls off and tends toward approximately flat. This critical frequency increases rapidly as the SRT region is traversed. Concomitantly, the integrated noise power $P$ markedly decreases.  Beyond the SRT region, where the magnetization is in-plane, $\nu_0$ exceeds the measurement bandwidth and $S(\nu)$ is small and nearly frequency-independent. Besides an overall scaling factor \cite{Crooker}, we emphasize that the spectral shape of $S(\nu)$ is not affected by the size of the laser spot. 

Empirically, we find that all these noise spectra can be reasonably fit over the measured frequency range to the functional form $S(\nu) \propto (\nu+\nu_{0})^{-3/2}$.  Examples of the fits are shown by dashed lines in Fig. 2a. Although not strictly physical [in principle, $S(\nu)$ should equal $S(-\nu)$], this fitting allows us to better identify how the total noise power $P$ and the critical frequency $\nu_0$ evolve through the SRT. Figure 2b shows how these parameters vary as a function of the lateral position $x$ across the sample. $P$ reaches a maximum at the location where the coercive field $\mu_0 H_c \rightarrow 0$. Near this point, $\nu_0$ begins to increase rapidly through the measurement frequency range and $P$ is subsequently rapidly suppressed.

As discussed in the previous section, the presence of $\nu_0$ is consistent with the development of a low-frequency truncation, or minimum rate $\gamma_{\rm min}$, in the distribution $D(\gamma)$ of relaxation rates exhibited by the Pt/Co/Pt film. The slowest out-of-plane magnetization fluctuations $\delta m_z(t)$, with characteristic relaxation rate less than $\nu_0$, are the first to be suppressed upon entering the SRT region.  As the SRT region is traversed and the magnetic anisotropy is further reduced, this low-frequency cutoff increases rapidly through our measurement bandwidth, after which point all out-of-plane fluctuations are very weak and the sample magnetization lies entirely in-plane.  These trends therefore provide a means of characterizing the loss of perpendicular magnetic order through the spin-reorientation transition by the frequency distribution of the thermodynamic fluctuations.

\subsection{Temperature-tuning through the SRT}
Because the SRT is defined by the interplay between magnetic anisotropy and thermal energy, we also can expect that the temperature $T$ will have a significant influence on the location of the SRT on the sample \cite{Pappas, Pescia, Jensen}, and therefore on the magnetization fluctuations measured at a fixed location.  This is confirmed in Fig. 3, where $S(\nu)$ is measured at a fixed location on the Pt/Co/Pt film as a function of $T$. Effectively, changing $T$ moves the SRT region through the location of the probe laser.  We find that $S(\nu)$ first exhibits a $\nu^{-3/2}$ power law across the measured 0.1~kHz to 200~kHz frequency range at low $T$ (where the magnetization is out-of-plane), then follows a $(\nu+\nu_{0})^{-3/2}$ dependence at intermediate $T$ (in the SRT regime), and finally evolves at high $T$ towards spectrally flat and small (where the magnetization has switched to in-plane). 

The effect of increased $T$ on $S(\nu)$ at a fixed location is therefore effectively equivalent to measuring different sample regions with increasing interfacial disorder at fixed $T$ (as in Fig. 2), and again we find that $\nu_0$ can be used to characterize the SRT.  It is also noteworthy that the exponent of the power-law decay, $\alpha = -3/2$, is insensitive to $T$ over the range of the experiment.  

\begin{figure}[tbp]
\center
\includegraphics[width=.45\textwidth]{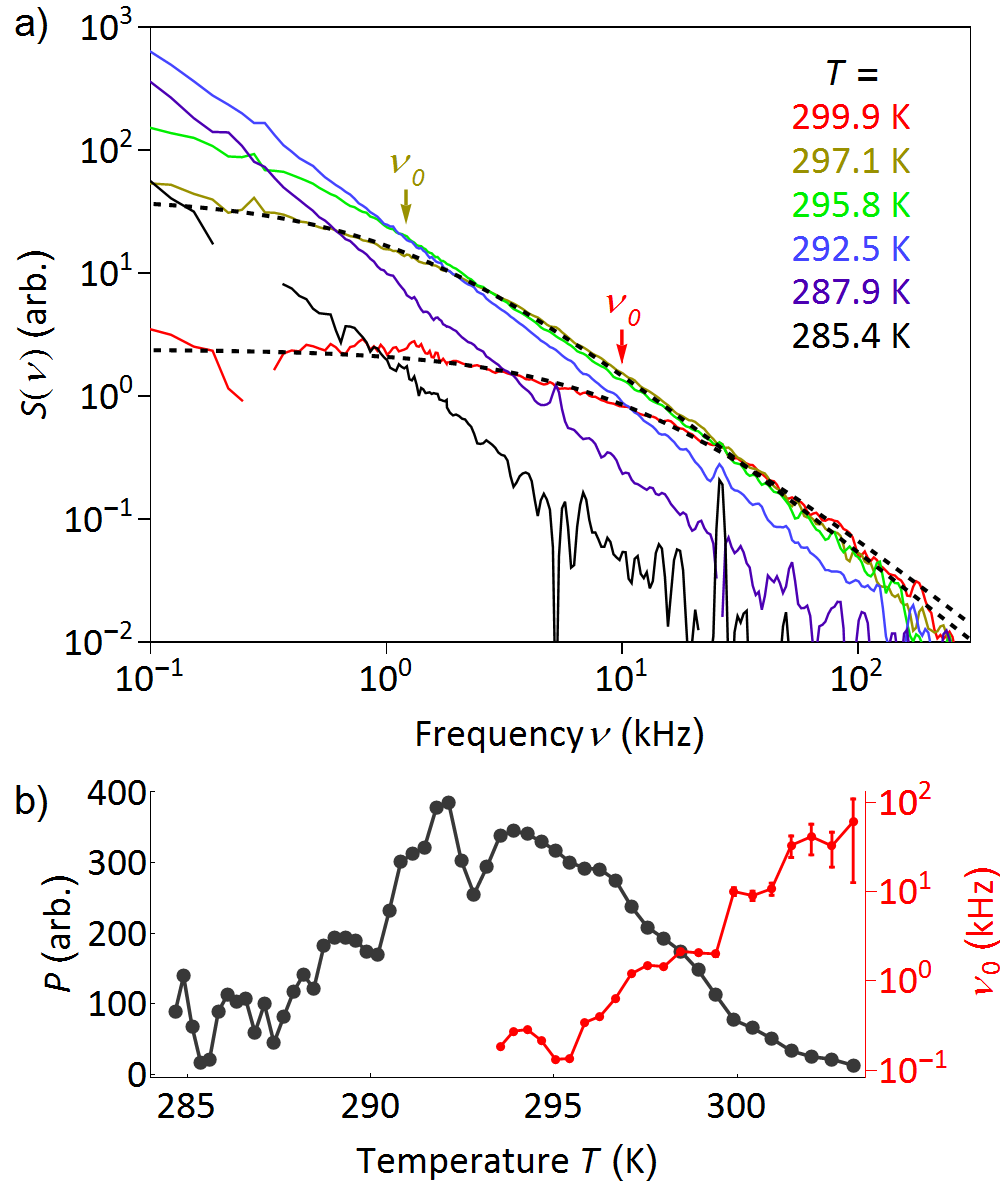}
\caption{a) Magnetization noise spectra $S(\nu)$ at different temperatures $T$, measured at a fixed location on the trilayer that exhibits perpendicular magnetization at low temperatures ($T<290$~K), and in-plane magnetization at high temperatures ($T>300$~K). $S(\nu)$ exhibits robust $\nu^{-3/2}$ power-law behavior at low $T$. With increasing $T$, $S(\nu)$ grows in magnitude but then develops a steadily-increasing critical frequency, $\nu_0$, below which the noise tends toward spectrally flat. Again, all spectra can be empirically fit by $S(\nu) \propto (\nu+\nu_{0})^{-3/2}$ (dashed lines).  b) Integrated noise power $P$ and critical frequency $\nu_0$ versus $T$, indicating that the SRT region has been temperature-tuned through the fixed location of the probe.}
\label{fig3}
\end{figure}

At other lateral positions on the sample with different net magnetic anisotropy, the SRT occurs at different temperatures.  Qualitatively similar thermodynamic fluctuations are nonetheless obtained (albeit peaked at different temperatures), as shown in Supplemental Fig. S1. In all cases studied, $S(\nu) \propto \nu^{-3/2}$ at low $T$, but develop an increasing cutoff frequency $\nu_0$ as $T$ is increased through the SRT. 

\subsection{Magnetization fluctuations in applied out-of-plane fields $B_z$, and relation to the susceptibility via the fluctuation-dissipation theorem}

Since the magnetization of these Pt/Co/Pt trilayers is strongly field-dependent, we next investigate how magnetization fluctuations $\delta m_z (t)$  are influenced by applied fields, beginning with the simplest case of purely out-of-plane fields $B_z$.  Here we expect that sufficiently large $|B_z|$ will completely align and saturate the magnetization $m_z$ along $\pm \hat{z}$, thereby suppressing fluctuations.  For reference, Fig. 4a shows the magnetization $m_z (B_z)$ at a location near the SRT, as measured by conventional polar MOKE using a very slow (quasi-static) field sweep rate of 22~$\mu$T/s. The magnetization exhibits no hysteresis and varies approximately linearly with $B_z$ (indicating a constant dc magnetic susceptibility), until it saturates when $|B_z| > B_{\rm sat}$, where $B_{\rm sat} \simeq 0.5$~mT. The absence of hysteresis indicates that at this location the trilayer can relax to an equilibrium maze-like magnetic domain pattern on timescales faster than the sweep rate of the external field.

Figure 4b shows a color surface map of the fluctuation spectra $S(\nu)$ versus $\nu$ and $B_z$, measured at the same location. Note that $S(\nu)$ and $\nu$ are both shown on logarithmic scales, so that equally-spaced contours indicate power-law decays. For all small applied fields $|B_z| < B_{\rm sat}$, $S(\nu)$ spectra are largely unaffected, and they decay with the same power-law exponent $\alpha \simeq -3/2$. However, $S(\nu) \rightarrow 0$ at all frequencies when $|B_z| > B_{\rm sat}$ and the magnetization saturates.

Figure 4c shows the integrated magnetization noise power, $P = \int S(\nu) d\nu$, as a function of $B_z$. Crucially, we note the close correspondence between $P$ (which is measured via magnetization fluctuation spectroscopy in Fig. 4b) and the quasi-dc magnetic susceptibility, $\chi = dm_z/dB_z$ (which is measured via conventional magnetometry in Fig. 4a). $P$ is large and approximately constant between $\pm B_{\rm sat}$, where $\chi$ is also large and approximately constant.  However, both $P$ and $\chi$ are rapidly suppressed to zero when $|B_z|>B_{\rm sat}$.  This correspondence is in line with the fluctuation-dissipation theorem \cite{Kubo}, which relates frequency-dependent magnetization fluctuations $S(\nu)$ to the frequency-dependent imaginary (\textit{i.e.}, dissipative) part of the magnetic susceptibility, $\chi'' (\nu)$:
\begin{equation}
\frac{2k_B T}{\pi \mu_0}\chi '' (\nu) = \nu S(\nu).
\end{equation}

\begin{figure}[tbp]
\center
\includegraphics[width=.43\textwidth]{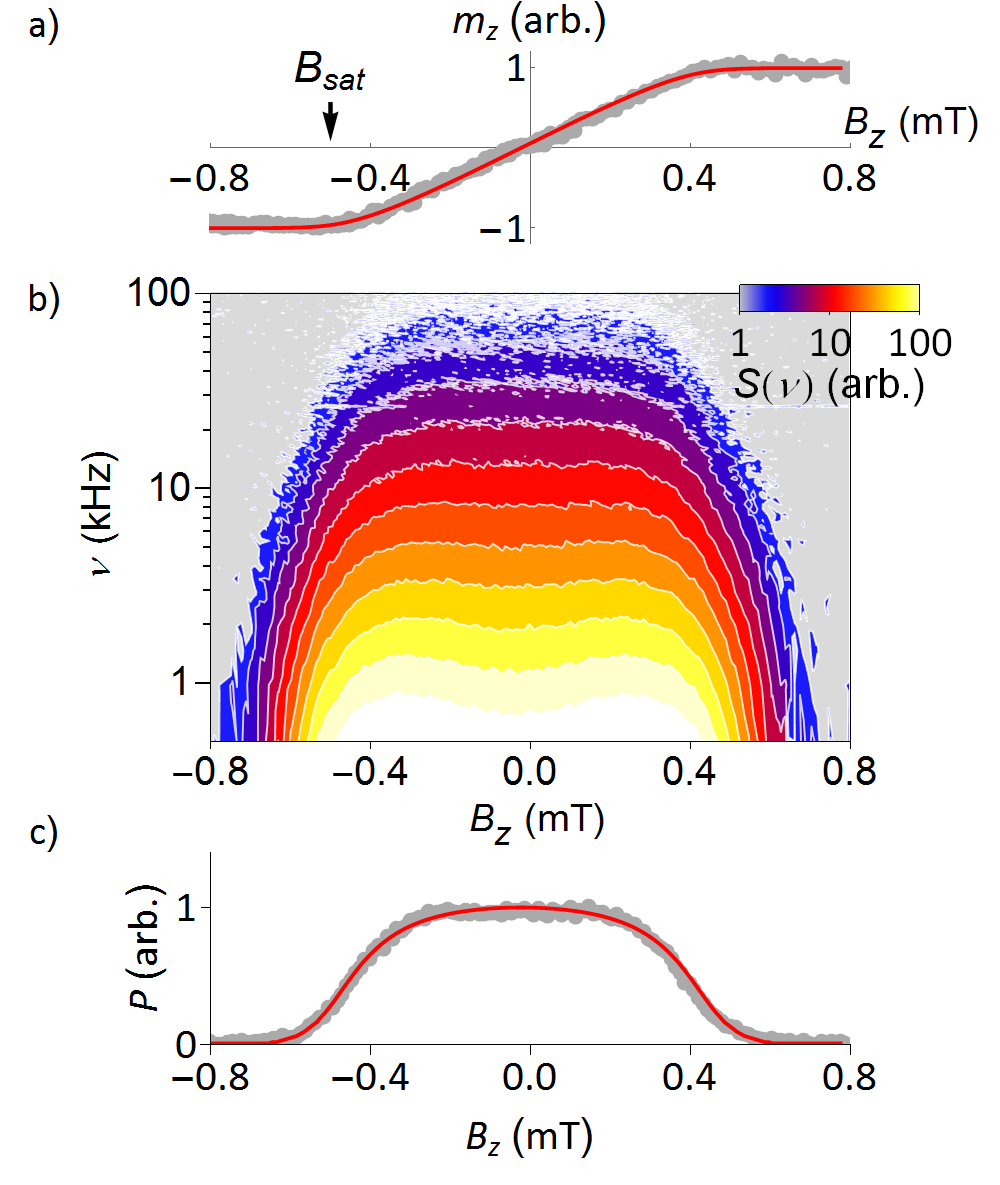}
\caption{a) The average magnetization $m_z$ versus applied perpendicular magnetic field $B_z$ (measured by conventional MOKE), at a location on the trilayer near the SRT. $m_z$ increases linearly with $B_z$ (indicating constant magnetic susceptibility $\chi = dm_z/dB_z$), until it saturates for $|B_z| > B_{\rm sat}$. b) Intensity map of the noise spectra $S(\nu)$ at different $B_z$. For all $|B_z|<B_{\rm sat}$, $S(\nu)$ is unaffected (and follows $\nu^{-3/2}$), but $S(\nu) \rightarrow 0$ when $|B_z| > B_{\rm sat}$. c) The integrated magnetization noise $P= \int S(\nu) d\nu$ versus $B_z$. The correspondence between $P$ and $\chi$ is in accord with the fluctuation-dissipation theorem.  Moreover, the detailed dependence of $P$ (and $m_z$) on $B_z$ can be qualitatively captured by simple considerations of the free energy of a mixture of magnetic domains, as shown by the red lines in panels (c) and (a) -- see text.}
\label{fig4}
\end{figure}

To relate $\chi'' (\nu)$ to the real dc magnetic susceptibility that was measured in Fig. 4a, we note that the real and imaginary parts of the susceptibility are necessarily coupled via Kramers-Kronig relations.  The dc ($\nu \simeq 0$) susceptibility is therefore given by
\begin{equation}
\chi (\nu \simeq 0) = \frac{2}{\pi}\int^{\infty}_{0} \frac{\chi ''(\nu ')}{\nu '} d\nu ' .
\end{equation}
Combining Eq. 2 and 3, we find that $\chi = dm_z/dB_z \propto \int S(\nu) d\nu = P$. Significant magnetization noise is therefore expected when the dc magnetic susceptibility is large \cite{Meier, Ocio, Diao}, in agreement with our measurements.

\subsection{Relating the magnetization noise to a simple model of the magnetic free energy}

We further note that the field dependence of both the average magnetization $m_z$ and the integrated magnetization noise $P$ can also be qualitatively captured by considering the free energy of a mixture of magnetic domains that are oriented along $\pm \hat{z}$ (i.e., within a domain the average magnetization $m_z=\pm 1$).  We use a minimal mean-field model to describe the free energy (per moment), $F=U-TS$.  It contains only a quadratic exchange term, a linear Zeeman energy term, and a typical Bragg-Williams mixing term for the entropy $S$:
\begin{equation}
\begin{split}
F = a m_z^2 - b B_z m_z - \frac{1}{2}k_B T [ 2{\rm ln} 2 - (1 + m_z) {\rm ln} (1 + m_z) -\\  (1 - m_z) {\rm ln} (1 - m_z)],
\end{split}
\end{equation}
where $a$ and $b$ are scaling constants. Since the data in Fig. 4a indicates that $m_z$=0 when $B_z$=0, $F$ has a minimum here and therefore we take the coefficient $a$ to be positive. $F(m_z)$ therefore has a single global minimum for any $B_z$. At any $B_z$, the equilibrium average magnetization $\langle m_z \rangle$ is given approximately by the value of $m_z$ for which $F$ is minimized (see Supplemental Fig. S2). More importantly, the curvature of $F(m_z)$ at this minimum value dictates the magnitude of fluctuations: if $F(m_z)$ increases only slightly for a given fluctuation $\langle m_z \rangle \pm \delta m_z$ (\textit{i.e.}, if the curvature $d^2F/dm_z^2$ is small), then fluctuations cost little energy and are thermodynamically likely. In other words the curvature of $F$ is inversely related to the variance of magnetization fluctuations, $(d^2F/dm_z^2)^{-1} \sim \langle (\delta m_z)^2 \rangle \sim P$ (for details see Supplemental Fig. S2).

The red lines in Figs. 4a and 4c show $m_z(B_z)$ and $P(B_z)$ calculated using Eq. 4, using a common set of coefficients $a$ and $b$ that are adjusted to match the experimental data. The minimum of $F(m_z)$ occurs at a value of $m_z$ that increases approximately linearly with $B_z$ until $m_z$ approaches $\pm 1$, and the curvature of $F(m_z)$ at this minimum is small and relatively constant until it increases sharply as $m_z \rightarrow \pm 1$. The overall trends are qualitatively captured, and again establish the link between the magnetization noise power $P$ and the dc magnetic susceptibility $\chi$.

\subsection{Mapping magnetization fluctuations in both out-of-plane and in-plane magnetic fields}

We also investigate how magnetization fluctuations in Pt/Co/Pt evolve under the additional influence of applied in-plane magnetic fields $B_x$.  In contrast to out-of-plane fields $B_z$, small in-plane fields are not expected to directly contribute to the Zeeman energy of perpendicularly-oriented magnetic domains, since $B_x \perp \hat{z}$.  However, in-plane fields \textit{can} influence the energy of the domain walls that bound the domains \cite{Thiaville, Je}. The change in domain wall energy at a particular location depends on the relative orientation of $B_x$ with respect to the local magnetization direction in the wall. The local energy can therefore increase or decrease. However, when integrating the \textit{total} domain wall energy around a closed path, linear changes in energy cancel out by symmetry and the total domain wall energy typically decreases quadratically as $B_x^2$ \cite{Thiaville, Je}. Besides influencing the energy (and fluctuations) of existing domain walls, application of $B_x$ at a point very near magnetic saturation can therefore also increase the likelihood of spontaneously forming new magnetic domains (such as bubble domains) \cite{Choi}.

\begin{figure}[tbp]
\center
\includegraphics[width=.42\textwidth]{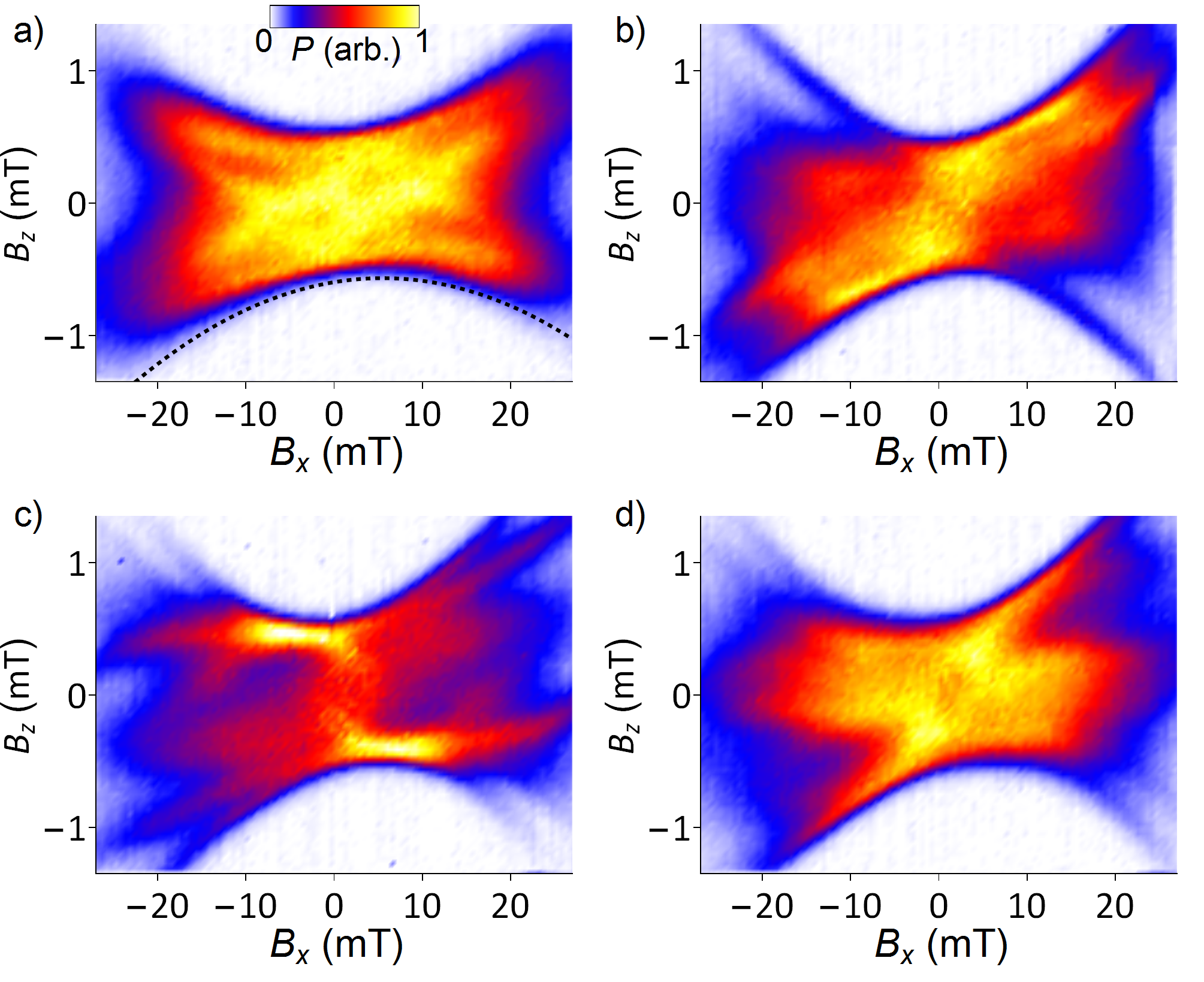}
\caption{a-d) Intensity maps of the integrated magnetization noise $P= \int S(\nu) d\nu$ versus in-plane and out-of-plane applied field ($B_x , B_z$). The four panels are acquired at slightly different locations on the Pt/Co/Pt film, separated by $<$10~$\mu$m. The bowtie shape of these images is determined by the balance between the fixed thermal energy $k_B T$ and the field-dependent energy required to spontaneously form magnetic domains (see text). The very different patterns within each bowtie, and the lack of reflection symmetry about the $B_{x,z}=0$ axes reveal a spatially-varying magnetic anisotropy that is canted slightly away from perfectly out of plane. Dashed line in a) shows a quadratic fit to the edge of the bowtie.}
\label{fig5}
\end{figure}

Figures 5a-d show intensity plots of the integrated magnetization noise power $P$ as a function of both $B_z$ (vertical axis) and also $B_x$ (horizontal axis), at four locations on the sample that are separated by only a few microns.  Rather striking ``bowtie'' structures are revealed (note, however, the factor-of-ten larger field scale for $B_x$).  A vertical line-cut through the bowtie at $B_x =0$ gives a cross-section that corresponds to the plot of $P(B_z)$ shown in Fig. 4c. The magnetic field at which $P$ falls abruptly to zero is therefore revealed by the upper and lower borders of the bowtie -- this is the applied field at which the average magnetization $m_z$ saturates along $\pm \hat{z}$, which was defined in Fig. 4 as $B_z = B_{\rm sat}$ for the case when $B_x$=0.

Several interesting properties of these fluctuation maps are immediately apparent. The first obvious aspect is that magnetization fluctuations persist out to larger $|B_z|$ when $\pm B_x$ is applied, and that this relationship is approximately quadratic in $B_x$, leading to the characteristic bowtie shape. This is consistent with the preceding discussion of how the net energy cost to form a closed domain wall decreases as $B_x^2$:  Consider, for example, a Pt/Co/Pt trilayer in an applied perpendicular field $B_z$ slightly larger than $B_{\rm sat}$, so that $m_z$ is saturated, the susceptibility $\chi = dm_z/dB_z$ is zero, and no magnetic domains are present. The available thermal energy $k_B T$ is not quite sufficient to induce the spontaneous formation of oppositely-oriented domains, and no fluctuations $\delta m_z(t)$ are present.  However, when $B_x$ is applied, the domain-formation energy falls below $k_B T$, at which point domains can spontaneously form and disappear, $m_z$ will no longer be saturated, $\chi$ will no longer equal zero, and magnetization fluctuations will appear. Only by increasing $B_z$ further (by an amount proportional to $B_x^2$) will the magnetization again achieve saturation.

A second obvious property of these images is that the total noise power $P$ is \textit{not} constant within a bowtie.  Patterns exist, but interestingly the patterns clearly lack reflection symmetry about the horizontal ($B_z$=0) and vertical axes ($B_x$=0). Rather, the images exhibit point inversion symmetry about the $B_{x,z}$=0 origin. This is consistent with a uniaxial magnetic anisotropy at the measurement location that is canted slightly \textit{away} from perfectly out-of-plane (\textit{i.e.}, it has some in-plane component), so that for a given $B_z$ the influence of $+B_x$ will in general be different than that of $-B_x$ (and similarly, at a given $B_x$ the influence of $+B_z$ and $-B_z$ will be different).  However, reversing both $B_x$ and $B_z$ preserves the symmetry and gives the same magnetic free energy, susceptibility, and noise, in keeping with the images in Fig. 5.

The final aspect that is quite apparent in Fig. 5 is that the four maps of $P(B_x, B_z)$ exhibit markedly different patterns, even though they were acquired at only slightly different locations on the Pt/Co/Pt trilayer ($<10$ $\mu$m apart). This indicates that deviations of the magnetic anisotropy away from $\pm \hat{z}$ are strongly position dependent in the Pt/Co/Pt film, at least down to length scales on the order of the laser spot (4~$\mu$m).   We emphasize that in order to acquire a map of $P(B_x, B_z)$, the sample is forced to undergo many saturation and demagnetization cycles; therefore the patterns cannot be due to some temporary or metastable domain configuration but must instead originate in a physical property of the sample. These variations in the magnetic anisotropy landscape likely derive from spatial inhomogeneity in the disorder that is induced by the Ar$^+$ irradiation and the fact that Co and Pt are not lattice matched.

\subsection{Inducing magnetization fluctuations in otherwise stable ferromagnetic films}
We now investigate how these noise maps evolve over a broad range of magnetic anisotropy.  Figure 6 shows maps of $P(B_x, B_z)$ acquired at very different lateral positions on the sample, starting  in panel (a) in the region with in-plane magnetization where fluctuations $\delta m_z(t)$ are small, continuing in panel (b) through the SRT region that was discussed in the previous section, and finally moving in panels (c-e) to less disordered regions of the sample with perpendicular magnetization and increasingly stable out-of-plane ferromagnetism.  For reference, the corresponding $m_z(B_z)$ magnetization curves (at $B_x$=0) are also shown. Of particular interest are Figs. 6 (c-e), which show how magnetization fluctuations evolve as the trilayer exhibits increasingly strong and stable perpendicular ferromagnetism. The bowties split apart, with fluctuations appearing at larger $|B_x|$.  As expected, no fluctuations are observed near zero field when the trilayer exhibits stable perpendicular ferromagnetism, which we define as the presence of an open hysteresis loop and a stable remnant magnetization. The key point, however, is that even in trilayers that are nominally very stable ferromagnets at zero field, magnetization fluctuations $\delta m_z(t)$ \textit{can} emerge in sufficiently large in-plane fields $B_x$.  Studies of non-irradiated trilayers with even larger $\mu_0 H_c$ confirm this to be generally true (see Supplemental Fig. S3).

\begin{figure}[tbp]
\center
\includegraphics[width=.46\textwidth]{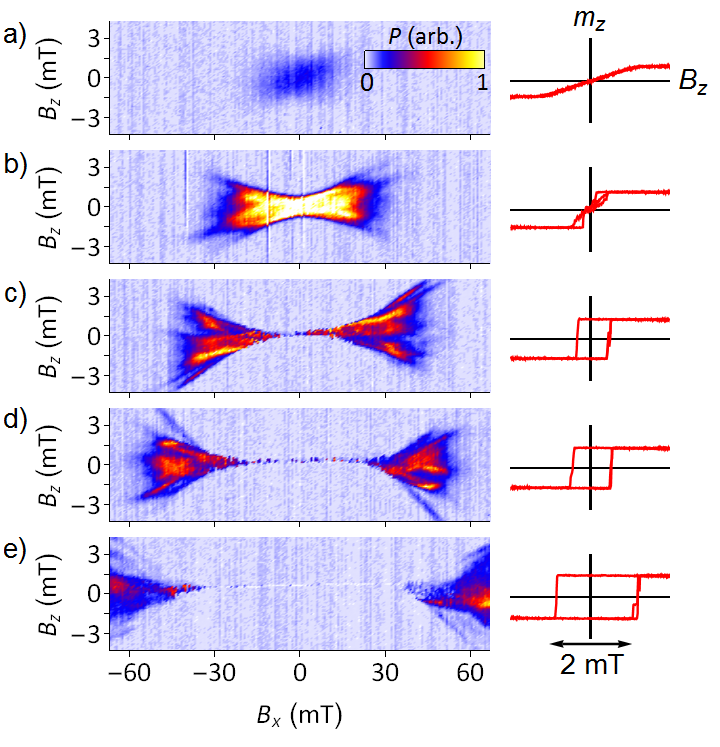}
\caption{Intensity maps of the integrated magnetization noise $P(B_x, B_z)$ at different locations across the Pt/Co/Pt sample, beginning in a) regions of the sample with in-plane magnetization, where $P$ is small. b) Near the SRT, \textit{P} is large when $|B_{z}| < B_{\rm sat}$ (\textit{cf.} Fig. 5). c-e) In regions with less disorder and therefore increasingly stable perpendicular magnetization, $\mu_0 H_c$ increases and fluctuations vanish at $B_{x,z}=0$.  However, fluctuations re-emerge when larger $|B_x |$ is applied.}
\label{fig6}
\end{figure}

This result has a number of consequences.  Similar to the data shown in Fig. 5, these noise maps show significant nonuniformity, indicating that even films with robust perpendicular ferromagnetism exhibit some degree of microstructured in-plane anisotropy, which could play a role, for example, during magnetic switching in technological applications. Further, it shows that measurements which use fluctuations to detect subtle micromagnetic effects \cite{Demidov} may be possible in materials with significant perpendicular anisotropy. Finally, we speculate that these results may be applicable to future neuromorphic magnetic computation schemes \cite{Mizrahi}, which require the capability to reversibly tune the fluctuations of individual magnetic elements and the interactions between them.

\subsection{Higher-order correlations of the magnetization noise}

In this last section, we demonstrate an important consequence and potential advantage of measuring fluctuations $\delta m_z(t)$ directly in the time domain, which is that \emph{all} possible time correlators can, in principle, be retrieved and analyzed from the noise signal. Thus far, we have focused  on $S(\nu)$, the power spectral density of the measured fluctuations, which is derived from the second-order (or two-point) time correlator $C_2(t)=\langle \delta m_z(0) \delta m_z(t)\rangle$. As previously described, $S(\nu) = \langle a(\nu) a^*(\nu) \rangle$ where $a(\nu) = (1/\sqrt{\tau}) \int_0^\tau \delta m_z(t) e^{i \nu t} dt$ is the Fourier transform of $\delta m_z(t)$ measured over a sufficiently long duration $\tau$.

Importantly, however, the $n^{th}$-order time correlator $C_n(t_1, t_2, ..., t_{n-1}) = \langle \delta m_z(0) \delta m_z(t_1)... \delta m_z(t_{n-1}) \rangle$ can contain additional information that is not trivially related to $S(\nu)$, particularly in the presence of interactions, inhomogeneous broadening effects, and/or non-gaussian noise due to (for example) the discrete nature of the system. In general, only the full set of all correlators contains complete information about an interacting system. Higher-order correlations have been studied theoretically \cite{Fuxiang, Schad} and experimentally in magnetic systems such as spin glasses \cite{WeissmanSG} and amorphous magnets \cite{Petta}, but we are not aware of any prior experimental studies of higher-order correlations of fluctuations in magnetic films in thermal equilibrium. Such correlators, however, play an important role in the theory of phase transitions and nonlinear thermodynamics \cite{Stratonovich}. Therefore their experimental measurement is highly desirable as a tool for characterization of material phases and for tests of fundamental theoretical predictions, such as higher-order fluctuation relations and for testing the universality of scaling exponents.  

By way of example we compute from our noise measurements the simplest nontrivial higher (third) order correlator $C_3 (t_1, t_2)$, which can be expressed in the frequency domain as 
\begin{equation}
C_3(\nu_1, \nu_2) = \langle a (\nu_{1}) a (\nu_{2}) a^* (\nu_{1}+\nu_{2}) \rangle.
\end{equation}
$C_3$ contains essentially different information compared to a two-point correlator.  By construction, it describes correlations between different frequencies and is therefore typically represented as a two-dimensional intensity plot.  Furthermore, note that $C_3$ is complex-valued.  Its real and imaginary components have different physical meaning, and in principle it is possible for fluctuations $\delta m_z(t)$ to exhibit a real $C_3$, or an imaginary $C_3$, or both.  The real part of $C_3$ is, for example, sensitive to any skewness or third moment in the distribution of $\delta m_z(t)$ about its mean -- e.g., if a simple magnetic fluctuator spends more time in its $+\hat{z}$ state than in its $-\hat{z}$ state (that being just one example of a non-gaussian noise distribution, as often exemplified by random telegraph noise with statistically inequivalent dwell times in the high and low state).  

\begin{figure}[tbp]
\center
\includegraphics[width=.46\textwidth]{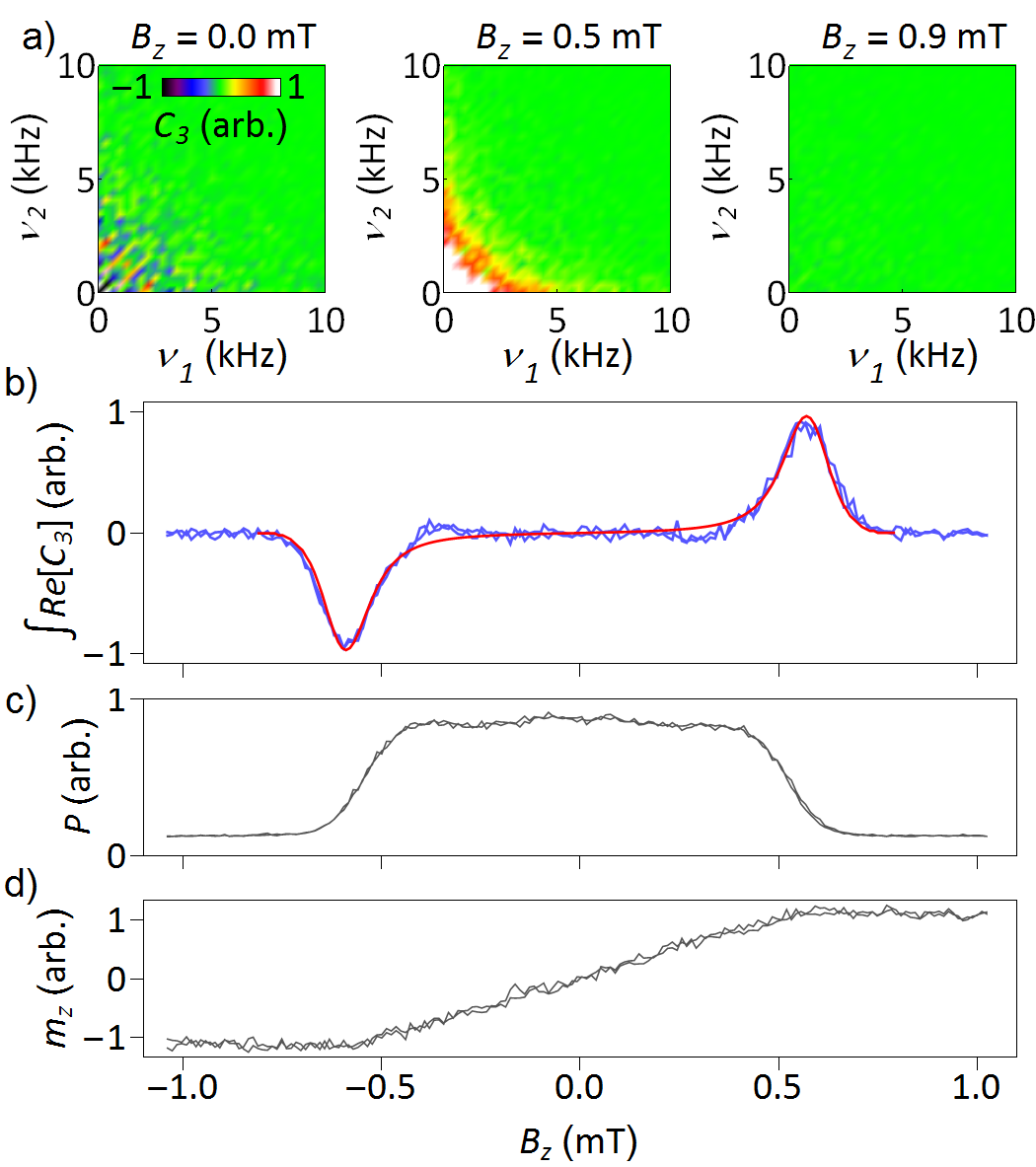}
\caption{Measuring higher-order correlations of the magnetization noise at a location on the sample near the SRT that exhibits linear magnetization and large fluctuations (similar to Fig. 4). a) The real part of the third-order correlator, ${\rm Re}[C_3(\nu_1, \nu_2)]$, acquired at $B_z$=0~mT, 0.5~mT, and 0.9~mT.  b) The integral $\int {\rm Re}[C_3] d\nu_1 d\nu_2$ measured as a continuous function of $B_z$. The red line shows this dependence calculated from the magnetic free energy (see Supplemental Material for details). c) The simultaneously-measured total noise power $P = \int S(\nu) d\nu$ versus $B_z$ (similar to Fig. 4c). d) The average magnetization $m_z(B_z)$ at this same location, as measured by conventional conventional MOKE, showing a constant susceptibility for $|B_z| < B_{\rm sat}$ (similar to Fig. 4a). The third-order correlator is approximately zero except when $|B_z| \simeq B_{\rm sat}$, indicating a skewness in the distribution of $\delta m_z(t)$.}
\label{fig7}
\end{figure}

The imaginary part of $C_3$ is sensitive to time-reversal symmetry breaking of the measured signal, meaning that $\delta m_z(t) \neq \delta m_z(-t)$ in a statistical sense. For example a sawtooth wave, though not a noise signal \textit{per se}, breaks time-reversal symmetry, but a triangular wave or sine wave does not. Random telegraph noise with statistically inequivalent rise and fall times would also generate an imaginary $C_3$, even when the dwell times in the high and low states are equivalent and the distribution of $\delta m_z(t)$ has no skewness. In general, magnetic systems can of course break time-reversal symmetry.  However, ${\rm Im}[C_3]=0$ is expected to hold for Ising-Glauber spin dynamics that satisfy detailed balance at thermal equilibrium \cite{Fuxiang}.

Figure 7 shows the real part of $C_3$, measured at a location on the trilayer where $m_z(B_z)$ increases linearly (and the noise power $S(\nu)$ is large) when $|B_z|<B_{\rm sat}$.  Fluctuations at this location are very similar to those shown earlier in Fig. 4, where $B_{\rm sat} \approx 0.5$~mT. Figure 7a shows ${\rm Re}[C_3(\nu_1, \nu_2)]$ measured at $B_z$=0~mT, 0.5~mT, and 0.9~mT.  Furthermore, Fig. 7b shows the integral $\int {\rm Re}[C_3] d\nu_1 d\nu_2$ as a continuous function of $B_z$.  For comparison and completeness, Figs. 7c and 7d show the simultaneously-measured total noise power $P = \int S(\nu) d\nu$ and the average magnetization $m_z$ as continuous functions of $B_z$ (\textit{cf.} Fig. 4).

The real part of $C_3$ is found to be non-zero when $B_z \approx B_{\rm sat}$ and $m_z$ approaches saturation. This indicates a measurable skewness (third moment) in the probability distribution of the noise signal $\delta m_z(t)$ about its mean, which can be viewed as arising from a cubic nonlinearity (i.e., a nonzero third derivative) of the free energy $F(m_z)$ at its minimum value. To show this, we note that integrating ${\rm Re}[C_3(\nu_1, \nu_2)]$ over both frequencies yields the equal-time correlator $\langle \delta m_z(t)^3 \rangle$, which in turn is proportional to $d^3F/dm_z^3$ (see Supplemental Fig. S2). By inspection of Eq. 4, this cubic nonlinearity arises from the entropic contribution to $F$ when $B_z \neq 0$. The data are in good qualitative agreement with predictions from Eq. 4, where ${\rm Re}[C_3]$ is small when $|B_z| << B_{\rm sat}$ (because the fluctuations are nearly gaussian) and when $|B_z| >> B_{\rm sat}$ (because the fluctuations are strongly suppressed), but reaches a local maximum when $|B_z| \approx B_{\rm sat}$ where both the cubic nonlinearity \textit{and} the fluctuations are large. The red line in Fig. 7b shows the real part of $C_3$ calculated from the magnetic free energy (derived in detail in the Supplementary Information), showing very good agreement.  We further note that other experimental gaussian noise sources such as photon shot noise do not produce a background level to $C_3$ that must be subtracted off (in contrast to $S(\nu)$), which further motivates analysis of $C_3$ to reveal subtle non-gaussianity effects in fluctuation signals.

Finally, we find that Im[$C_3 (\nu_1, \nu_2)$]=0 to within our experimental accuracy, for all applied $B_z$. This means that the measured fluctuation signal $\delta m_z(t)$ in our Pt/Co/Pt trilayers did not explicitly break time-reversal symmetry, and that in thermal equilibrium $\delta m_z(t)$ remained indistinguishable from $\delta m_z(-t)$ in a statistical sense. 

\subsection{Summary}

We have used Kerr magnetometry to study the broadband frequency spectra of thermodynamic magnetization fluctuations in an archetypal ultrathin ferromagnet (a Pt/Co/Pt trilayer).  The power spectral density of the fluctuations, $S(\nu)$, is found to follow a robust $\nu^{-3/2}$ power law on regions of the trilayer exhibiting out-of-plane magnetization, indicating a broad distribution of fluctuation and relaxation rates.  However, the functional form of $S(\nu)$ changes dramatically as the magnetic anisotropy is reduced to zero and the trilayer is tuned through a SRT (either by increasing temperature or by increasing interfacial disorder).  Namely, as the SRT is traversed, $S(\nu)$ develops a steadily-increasing critical frequency $\nu_0$ below which the noise power tends towards spectrally flat, indicating an evolving low-frequency cutoff for out-of-plane magnetization fluctuations.  These results therefore provide a means of characterizing phase transitions in ferromagnets via their influence on spontaneous dynamics. The fluctuations are found to be strongly dependent on applied magnetic fields, which can be understood within the framework of the fluctuation-dissipation theorem and from considerations of the magnetic free energy. Sufficiently large in-plane fields can induce spontaneous fluctuations in films that otherwise exhibit stable out-of-plane ferromagnetism. Finally, we demonstrate that higher-than-second order correlators can be computed and analyzed from the measured fluctuations, which can potentially provide a sensitive and powerful means of revealing subtle influence of non-gaussian noise and time-reversal breaking effects.  

We thank Dan Pierce, Vivien Zapf, Mark Stiles, Alice Mizrahi, and Carl Boone for useful discussions. Work at the NHMFL was supported by the Los Alamos LDRD program, National Science Foundation DMR-1157490, the State of Florida, and the US Department of Energy.

\end{document}